\documentclass[12pt]{article}
\pdfoutput=1
\DeclareMathAlphabet{\scr}{U}{rsfs}{m}{n}
\usepackage{latexsym}
\usepackage[mathscr]{eucal}
\usepackage{amsfonts}
\usepackage{amscd}
\usepackage{cite}
\usepackage{amssymb}
\usepackage[centertags]{amsmath}
\usepackage{dsfont}
\usepackage{yfonts}
\usepackage{enumerate}
\usepackage{graphicx}
\usepackage{feynmf}

\newcommand{\cleqn}{\setcounter{equation}{0}}
\newcommand{\newc}{\newcommand}
\newc{\be}{\begin{equation}}
\newc{\ee}{\end{equation}}
\newc{\bea}{\begin{eqnarray}}
\newc{\eea}{\end{eqnarray}}
\newc{\ben}{\begin{equation*}}
\newc{\een}{\end{equation*}}
\newc{\bean}{\begin{eqnarray*}}
\newc{\eean}{\end{eqnarray*}}
\newc{\ol}{\overline}
\newc{\wt}{\widetilde}
\newc{\bs}{\boldsymbol}
\newc{\m}{\mathcal}
\newc{\la}{\lambda}
\newc{\lra}{\longrightarrow}
\newc{\vp}{\varphi}
\newc{\ti}{\tilde}

\begin{document}

\title{\hfill ~\\[-30mm]
       \hfill\mbox{\small \begin{tabular}{l}
                                   DFPD-12/TH/2\\[1mm]
                                   IPPP-12-27\\[1mm]
                                   DCPT-12-54\\[1mm]
                          \end{tabular}}\\[15mm]
       \textbf{SUSY $\bs{S_4 \times SU(5)}$ revisited }}
\date{}

\author{\\Claudia Hagedorn\footnote{E-mail: {\tt hagedorn@pd.infn.it}}$\;\;^a$, Stephen F. King\footnote{E-mail: {\tt king@soton.ac.uk}}$\;\;^b$,
Christoph Luhn\footnote{E-mail: {\tt  christoph.luhn@durham.ac.uk}}$\;\;^c$\\[10mm]
$^a$ \emph{\small{}Dipartimento di Fisica e Astronomia `G.~Galilei', Universit\`a di Padova}\\
  \emph{\small INFN, Sezione di Padova,}\\
  \emph{\small Via Marzolo~8, I-35131 Padua, Italy}\\[2mm]
 $^b$ \emph{\small{}School of Physics and Astronomy, University of Southampton,}\\
  \emph{\small Southampton, SO17 1BJ, United Kingdom}\\[2mm]
 $^c$ \emph{\small Institute for Particle Physics Phenomenology, University of Durham,}\\
  \emph{\small Durham, DH1 3LE, United Kingdom}}

\maketitle

\begin{abstract}
\noindent 

Following the recent results from Daya Bay and RENO, which measure the lepton
mixing angle $\theta^l_{13} \approx 0.15$, we revisit a supersymmetric (SUSY)
$S_4 \times SU(5)$ model, which predicts  tri-bimaximal (TB) mixing in the neutrino sector 
with $\theta_{13}^l$ being too small in its original version. We show that
introducing one additional $S_4$ singlet flavon into the model gives rise to
a sizable $\theta^l_{13}$ via an operator  which leads to the breaking of one
of the two $Z_2$ symmetries preserved in the neutrino sector at leading order (LO).
The results of the original model for fermion masses, quark mixing and the
solar mixing angle are maintained to good precision. The atmospheric and solar
mixing angle deviations from TB mixing are subject to simple sum rule bounds.

\end{abstract}
\thispagestyle{empty}
\vfill
\newpage
\setcounter{page}{1}


\section{Introduction}
\cleqn


Global fits have indicated a non-zero value for the lepton mixing angle
$\theta_{13}^l$ for some time \cite{firsttheta13nonzero}. Direct evidence for
a large $\theta^l_{13}$ has been provided in 2011 by T2K \cite{T2K}, MINOS
\cite{MINOS} and Double Chooz \cite{DC};  for global fits including 
the results of T2K and MINOS see \cite{fogli,schwetz,maltoni}. Recently, Daya
Bay \cite{DayaBay} have published  their first result
\be
\sin^2 2  \theta^l_{13}=0.092\pm 0.016 \, \mathrm{(stat.)} \pm 0.005 \, \mathrm{(syst.)},
\ee
while 
RENO \cite{RENO} find 
\be
\sin^2 2  \theta^l_{13}=0.113\pm 0.013 \, \mathrm{(stat.)} \pm 0.019 \, \mathrm{(syst.)},
\ee
implying $\theta^l_{13}\approx 0.15 \div 0.17$. Many models predicting TB
mixing \cite{HPS} at LO, in particular those with the flavour symmetries $A_4$
and $S_4$, for reviews see \cite{reviews}, are severely challenged by such a
large value of $\theta_{13}^l$,  because subleading corrections are too small
to explain $\theta^l_{13}\sim 0.15$.  

In this note we revisit an existing SUSY $S_4 \times SU(5)$ model
\cite{HKL10}. In its original form, it predicts TB mixing in the neutrino
sector,  and the dominant source of the mixing angle $\theta_{13}^l$ is the (12)-rotation in
the charged lepton sector which in turn is related to the Cabibbo angle
$\theta_C$ so that $\theta^l_{13} \approx \theta_C/(3 \sqrt{2}) \approx 0.05$.   
We show that a minimal extension of this model, namely adding one flavon
$\eta$, being a singlet under $S_4$, naturally gives rise to a value of
$\theta^l_{13}$ as indicated by the latest experimental results. An operator
is induced with the help of the field $\eta$ which gives rise to a
contribution to the neutrino mass matrix that breaks one of the two $Z_2$
symmetries responsible for TB mixing in the neutrino sector,  while
tri-maximal (neutrino) mixing is still protected by the intact $Z_2$ symmetry.  
The broken $Z_2$ symmetry is  identified with $\mu -\tau$ symmetry,
 and thus it becomes possible to generate $\theta_{13}^l$ of the correct
  size. At the same time, one encounters a deviation of the atmospheric mixing
  angle from its maximal value which is proportional to $\theta_{13}^l$.
The introduction of $\eta$ does not affect the successful predictions of the
original $S_4 \times SU(5)$ model. In particular, the Gatto-Sartori-Tonin
(GST) relation \cite{GST}, the size of the quark mixing angles $\theta_{13}^q$
and $\theta_{23}^q$, and the Georgi-Jarlskog (GJ) relations \cite{GJ} are
unaltered. Also the leading correction to the solar mixing angle remains the
same, whereas the atmospheric mixing angle, as explained, receives larger
corrections. As will be shown, the deviations of both these angles from TB
mixing are subject to sum rule bounds. Furthermore, we discuss a simple
ultraviolet (UV) completion of the operators  directly relevant for fermion
masses and mixing in this model. Note that the way of generating large
$\theta_{13}^l$ discussed here is very similar to the one proposed in
\cite{KL11} where it has been realised in the context of a non-unified SUSY $S_4$ model. 

The structure of the paper is as follows: in section \ref{recap} we
recapitulate the main features and results of the original model; in section
\ref{sec3} we discuss the operators which arise from introducing the new
flavon $\eta$ and show the results for the lepton mixing; section
\ref{flavons} contains a brief discussion of the relevant parts of the flavon
superpotential, while we present the messengers of a simple UV completion
in section \ref{messengers}. We comment on the differences between the choices
$\eta \sim {\bf 1}$ and $\eta \sim {\bf 1'}$ under $S_4$ in section
\ref{eta1pr} and conclude in section \ref{conclusions}.


\section{Original SUSY $\bs{S_4 \times SU(5)}$ model}
\label{recap}
\cleqn


We consider a SUSY $SU(5)$ model in four dimensions. 
The full flavour symmetry of the model is $S_4 \times U(1)$. For details of
the group theory of $S_4$ we refer to the appendix of \cite{HKL10}. The $U(1)$
symmetry is necessary to control the operators with several
flavons.\footnote{For simplicity, we assume it to be a global symmetry. For a
  discussion on its nature see \cite{HKL10}.}  The three generations of ${\bf
  10}$-plets, $T$ and $T_3$, transform as ${\bf 2} + {\bf 1}$ under $S_4$, 
while the ${\bf \bar{5}}$-plets $F$ and the three gauge singlets $N$ form
triplets ${\bf 3}$ under $S_4$. At LO, two flavon doublets $\Phi^u_2$ and
$\wt\Phi^u_2$ are responsible for up quark masses, while three flavon
multiplets $\Phi^d_2 \sim {\bf 2}$, $\Phi^d_3 \sim {\bf 3}$, $\wt\Phi^d_3 \sim
{\bf 3}$ are necessary for generating down quark and charged lepton masses as
well as the quark mixing angles.  Right-handed neutrinos become massive when the
three flavon multiplets $\Phi^\nu_1 \sim {\bf 1}$, $\Phi^\nu_2 \sim {\bf 2}$
and $\Phi^\nu_{3'} \sim {\bf 3'}$ acquire non-vanishing vacuum expectation values (VEVs).
The specific alignment of the latter preserves a $Z_2 \times Z_2$ subgroup of
the group $S_4$ and leads to the prediction of TB mixing in the neutrino
sector. The model contains three $SU(5)$ Higgs fields $H_{5}$, $H_{\ol{5}}$
and $H_{\ol{45}}$ being singlets of $S_4$. With the correct vacuum, the latter
is responsible for the GJ relations. Although the model is formulated at the
scale of grand unification, we shall not be concerned with the details of
$SU(5)$ breaking which we assume to be unrelated to flavour physics. The particle content
and its transformation properties under $SU(5) \times S_4 \times U(1)$ are
summarised in table \ref{particles} together with the new field $\eta$ which
we assume to transform as a trivial singlet under $S_4$. Note that we fix the
$U(1)$ charges in the present note to be $(x,y,z)=(5,4,1)$, since it has been
shown in \cite{HKL10} that a model with such charges does not suffer from
operators which significantly perturb the LO results of fermion masses and
mixing and/or the vacuum alignment. 
\begin{table}
\begin{center}
$$
\begin{array}{|c||c|c|c|c|c|c|c|c|c|c|c|c|c|c|c||c|}\hline
\text{Field}\!\!\phantom{\Big|} & T_3 & T & F & N & H^{}_{5} & H_{\ol{5}} & H_{\ol{45}} &  \Phi^u_2 & \wt\Phi^u_2 & \Phi^d_3 & \wt\Phi^d_3 & \Phi^d_2  & \Phi^\nu_{3'} & \Phi^\nu_2 & \Phi^\nu_1 & \eta\\\hline
\!SU(5)\!\!\!\phantom{\Big|} & \bf 10 & \bf 10 & \bf \ol 5 & \bf 1 &\bf  5 &\bf \ol 5 &\bf \ol{45}
&\bf 1&\bf 1&\bf 1&\bf 1&\bf 1&\bf 1&\bf 1&\bf 1& \bf 1\\\hline
S_4\!\!\phantom{\Big|} & \bf 1&\bf 2&\bf 3&\bf 3&\bf 1&\bf 1&\bf 1&\bf 2&\bf 2&\bf 3&\bf 3&\bf
2&\bf 3'&\bf 2&\bf 1 & \bf 1\\ \hline
U(1)\!\!\phantom{\Big|} & 0&5&4&-4&0&0&1&\!-10\!&0&\!-4\!&\!-11\! &1 &8&8&8& 7 \\\hline
\end{array}
$$
\end{center}
\caption{\label{particles} The symmetries and charges of the superfields in
  the $SU(5)\times S_4 \times U(1)$ model. Compared to table 1 in
  \cite{HKL10}, we fix the $U(1)$ charges $(x,y,z)=(5,4,1)$ and have added the field $\eta$.}  
\end{table}
The leading operators of the matter superpotential are (order one coefficients
are suppressed with the exception of the terms involving right-handed neutrinos $N$) 
\bea
\label{LOup}
&& T_3T_3H^{}_{5} + \frac{1}{M} T T  \Phi^u_2 H^{}_{5}  +  \frac{1}{M^2} TT \Phi^u_2 \wt\Phi^u_2 H^{}_{5} 
\\ \label{LOdown}
&&+ \frac{1}{M} F T_3 \Phi^d_3 H_{\ol{5}} + \frac{1}{M^2} (F \wt\Phi^d_3)_{\bf{1}} ( T \Phi^d_2 )_{\bf{1}} H_{\ol{45}}
+ \frac{1}{M^3} (F \Phi^d_2 \Phi^d_2)_{\bf{3}} ( T \wt\Phi^d_3 )_{\bf{3}} H_{\ol{5}}
\\ \label{LOnu}
&&+ y_D F N H^{}_{5} + \alpha N N \Phi^\nu_1 + \beta N N \Phi^\nu_2 + \gamma N N \Phi^\nu_{3'} \, ,
\eea
where $M$ is the generic messenger mass which is of the order of the scale of grand unification.
Note that we have to assume specific contractions, ${\bf 1}$ and ${\bf 3}$, to
dominate in the case of the second and third operators in Eq.(\ref{LOdown}),
as has been discussed in detail in \cite{HKL10}. A viable UV completion
leading to the scenario in which these operators are dominant is briefly
recapitulated in section \ref{messengers}. The following vacuum alignment,
achieved through $F$-terms of suitable driving fields \cite{HKL10},  
\bea
\label{VEVup}
&& \langle \Phi^u_2 \rangle ~ = ~ \varphi^u_2 \begin{pmatrix} 0 \\ 1 \end{pmatrix} \; , \;\;
\langle \wt\Phi^u_2 \rangle ~=~
\wt\varphi^u_2 \begin{pmatrix}0\\1\end{pmatrix} \; , \;\;
\\ \label{VEVdown}
&& \langle \Phi^d_3 \rangle ~ = ~ \varphi^d_3 \begin{pmatrix}0\\1\\0\end{pmatrix} \; , \;\;
\langle \wt\Phi^d_3 \rangle ~ =~
\wt\varphi^d_3 \begin{pmatrix}0\\-1\\1\end{pmatrix} \; , \;\;
\langle \Phi^d_2 \rangle ~=~ \varphi^d_2 \begin{pmatrix}1\\0\end{pmatrix} \; , \;\;
\\ \label{VEVnu}
&& \langle\Phi^\nu_{3'} \rangle~=~ \varphi^\nu_{3'} \begin{pmatrix}1\\1\\1 \end{pmatrix} \; , \;\;
\langle\Phi^\nu_2\rangle~=~ \varphi^\nu_2 \begin{pmatrix}1\\1 \end{pmatrix} \; , \;\;
\langle \Phi^\nu_1 \rangle~=~ \varphi^\nu_1 \; ,
\eea
leads to a diagonal up quark mass matrix $M_u$, a down quark mass matrix $M_d$
and a charged lepton mass matrix $M_e$ of the form 
\be
\label{matrixMd}
M_d \approx
\begin{pmatrix}
0 & (\varphi^d_2)^2 \wt\varphi^d_3/M^3  & -(\varphi^d_2)^2 \wt\varphi^d_3/M^3   \\
-(\varphi^d_2)^2 \wt\varphi^d_3/M^3   & \varphi^d_2 \wt\varphi^d_3/M^2 &   -\varphi^d_2 \wt\varphi^d_3/M^2  + (\varphi^d_2)^2 \wt\varphi^d_3/M^3   \\
0  & 0 & \varphi^d_3/M
\end{pmatrix} v_d 
\ee
and
\be
\label{matrixMe}
M_e \approx
\begin{pmatrix}
0 & -(\varphi^d_2)^2 \wt\varphi^d_3/M^3  & 0\\
(\varphi^d_2)^2 \wt\varphi^d_3/M^3   & -3 \, \varphi^d_2 \wt\varphi^d_3/M^2  & 0\\
-(\varphi^d_2)^2 \wt\varphi^d_3/M^3   & 3 \, \varphi^d_2 \wt\varphi^d_3/M^2  +
(\varphi^d_2)^2 \wt\varphi^d_3/M^3  & \varphi^d_3/M
\end{pmatrix} v_d 
\ee
(with $v_d$ being the VEV of the light  combination of electroweak doublets
contained in $H_{\ol{5}}$ and $H_{\ol{45}}$). The Dirac neutrino mass matrix 
and the right-handed neutrino mass matrix read
\be
M_D = y_D \begin{pmatrix}
                     1 & 0 & 0\\
		     0 & 0 & 1\\
		     0 & 1 & 0
\end{pmatrix} v_u 
\;\; , \;\;\;
\label{matrixMR}
M_R = \begin{pmatrix}
                     \alpha \varphi^\nu_1 + 2 \gamma \varphi^\nu_{3'} & \beta \varphi^\nu_2 - \gamma \varphi^\nu_{3'}
		              & \beta \varphi^\nu_2 -\gamma \varphi^\nu_{3'}\\
		      \beta \varphi^\nu_2 - \gamma \varphi^\nu_{3'} & \beta \varphi^\nu_2 + 2 \gamma \varphi^\nu_{3'}
		              & \alpha \varphi^\nu_1 - \gamma \varphi^\nu_{3'}\\
		     \beta \varphi^\nu_2 -\gamma \varphi^\nu_{3'} &  \alpha \varphi^\nu_1 - \gamma \varphi^\nu_{3'}
		              & \beta \varphi^\nu_2 + 2 \gamma \varphi^\nu_{3'}
\end{pmatrix} 
\ee
($v_u$ is the VEV of the electroweak doublet contained in $H_{5}$). The matrix $M_R$
is of the most general form compatible with TB mixing. For $\varphi^u_2$,
$\wt\varphi^u_2 \sim \la^4 M$,   $\la \approx \theta_C \approx
0.22$, the up and charm quark mass, $m_u \approx \varphi^u_2 \wt\varphi^u_2
v_u/M^2$ and $m_c \approx \varphi^u_2 v_u/M$, are correctly produced, while
the top quark mass is of order $v_u$ being generated through a renormalisable
operator, see Eq.(\ref{LOup}). For small and moderate values of $\tan\beta$, $\tan \beta = v_u/v_d$,
$\varphi^d_3 \sim \la^2 M$ gives rise to the correct bottom quark and tau
lepton mass. The Cabibbo angle requires $\varphi^d_2 \sim \la \, M$ so that
$\theta_C \approx \tilde{x}_2/y_s \la$, see \cite{HKL10} for details. 
The correct size of the mass of the strange quark and of the muon is achieved
for $\wt\varphi^d_3 \sim \la^3 M$. As one can check, the electron and the down
quark mass are also of the correct order of magnitude and the masses of
charged leptons and down quarks fulfil the GJ relations
\cite{GJ}. Furthermore, the GST relation holds in the quark sector 
\cite{GST}. Eventually, the VEVs $\varphi^\nu_{1,2,3'}$ of the flavons
$\Phi^\nu_1$, $\Phi^\nu_2$ and $\Phi^\nu_{3'}$ are of order $\la^4 M$ in order
to correctly generate the light neutrino mass scale $m_\nu \sim 0.1 \,
\mathrm{eV}$. The light neutrino mass spectrum can have either hierarchy.
In the original model the value of the lepton mixing angle $\theta_{13}^l$ is
determined by the mixing in the charged lepton sector \cite{th13troughth12e}
which can be read off from the matrix $M_e$ in Eq.(\ref{matrixMe}),
$\theta_{12}^e \approx \theta_C/3 \approx \la/3$:  
\be
\sin \theta_{13}^l \approx \theta^e_{12} \sin\theta^\nu_{23} \approx \theta^e_{12}/\sqrt{2} \approx \la/(3 \sqrt{2}) \approx 0.05 \, ,
\ee
for TB mixing in the neutrino sector. The other two lepton mixing angles are
\be
\sin^2 \theta^l_{23} \approx 1/2 \; , \;\; \sin^2 \theta^l_{12} \approx 1/3 +
2/9 \, \la \cos \delta^l \ ,
\ee
with $\delta^l$ being the leptonic Dirac CP phase. As has been shown in detail
in \cite{HKL10} subleading corrections coming from operators with several
flavons hardly alter the fermion mass matrices and the vacuum alignment,
presented in Eqs.(\ref{VEVup}-\ref{VEVnu}). Thus all results achieved at LO
are uncorrected to good approximation. 


\section{Extended SUSY $\bs{S_4 \times SU(5)}$ model}
\label{sec3}
\cleqn


\subsection{Generation of large $\bs{\theta_{13}^l}$}
\label{largeth13}

A simple way to enhance the value of $\theta_{13}^l$ is to add a new flavon
$\eta$ which couples to right-handed neutrinos and gives rise to a
contribution which partly breaks the $Z_2 \times Z_2$ subgroup preserved by
the LO VEVs of the flavons $\Phi^\nu_{1}$, $\Phi^\nu_{2}$ and
$\Phi^\nu_{3'}$. Recall that the $Z_2 \times Z_2$ symmetry 
is generated by $S$ and $U$, defined in the appendix of \cite{HKL10}, and is responsible for
TB mixing in the neutrino sector. Adding a field $\eta$ which transforms as
${\bf 1}$ under $S_4$ and carries $U(1)$ charge $+7$, see table
\ref{particles}, allows to write down the term 
\be
\label{theta13op}
\eta \Phi^d_2 N N/M \, .
\ee
The VEV of $\Phi^d_2$,  see Eq.(\ref{VEVdown}), breaks the $Z_2$ symmetry
generated by  
\be
U = \left( \begin{array}{cc}
0 & 1\\
1 & 0
\end{array}
\right) 
\ee
in the chosen basis, because it is not an eigenvector of $U$. On the other
hand, it trivially leaves invariant the $Z_2$ symmetry generated by $S$, since
$S=\mathds{1}$ for the representation ${\bf 2}$. In this way a non-zero value
of $\theta_{13}^l$ is generated in the neutrino sector, while  tri-maximal
mixing related to the generator $S$ is still maintained, see Eqs.(\ref{mnuLO},
\ref{Strimax}).\footnote{Tri-maximal mixing has been discussed in
  \cite{trimax}.} Knowing that $U$ is represented for ${\bf 3}$ and ${\bf 3'}$ by
\be
U = \mp \left( \begin{array}{ccc}
1 & 0 & 0\\
0 & 0 & 1\\
0 & 1 & 0
\end{array}
\right) \, ,
\ee
one immediately sees that the broken $Z_2$ symmetry is identified with $\mu
-\tau$ symmetry which protects $\theta_{13}^\nu$ as well as $\theta_{23}^\nu$
from deviating from their TB mixing values. As a consequence, we expect that
the atmospheric mixing angle is corrected in a similar way to
$\theta_{13}^\nu$. For $\theta_{13}^\nu \sim \la$ this implies a correction
term proportional to $\la$ also for the atmospheric mixing angle. This
expectation is confirmed by Eqs.(\ref{thetanu},\ref{asr}). 
The requirement that $\theta_{13}^l \sim \la$ originates from the
operator in Eq.(\ref{theta13op}) implies that the order of the VEV of $\eta$ has to be 
\be
\langle\eta\rangle \approx \la^4 \, M \, ,
\ee
since the leading terms giving rise to right-handed neutrino masses are of the
order $\la^4 \, M$, see Eq.(\ref{matrixMR}). The explicit form of the
contribution due to the term in Eq.(\ref{theta13op}) is  
\be
\langle\eta\rangle \varphi^d_2 \, \left(N_1 N_2 + N_2 N_1 + N_3 N_3 \right)/M
\, ,
\ee
using the LO vacuum in Eq.(\ref{VEVdown}). Note that we can also write down an
operator similar to the one in Eq.(\ref{theta13op}) for the choice $\eta\sim
{\bf 1'}$. We comment on this possibility and the differences between the two
choices, $\eta \sim {\bf 1}$ and $\eta \sim {\bf 1'}$, in section
\ref{eta1pr}. The light neutrino mass matrix, arising from the type I see-saw
mechanism, can be cast into the form 
\be
\label{mnuLO}
m_\nu^{eff} =\left( \begin{array}{ccc}
                b_\nu +c_\nu-a_\nu & a_\nu & a_\nu + d_\nu \, \la \\
                a_\nu  & b_\nu + d_\nu \, \la & c_\nu \\
	 a_\nu + d_\nu \, \la & c_\nu & b_\nu
\end{array}
\right) \, \left( \frac{v_u^2}{\la^4 M} \right) \, .
\ee
The matrix  $m^{eff}_\nu$ clearly shows that the tri-maximal vector
$(1,1,1)^T$ is still an eigenvector of $m_\nu^{eff}$, even for $d_\nu \neq 0$,
which is traced back to the invariance of the neutrino mass matrix under $S$ 
\be
\label{Strimax}
S^T \, m_\nu^{eff} \, S= m_\nu^{eff} \;\;\;\;\; \mbox{with} \;\;\;\;\;
S= \frac{1}{3} \, \left( \begin{array}{ccc}
-1 & 2 & 2\\
2 & -1 & 2\\
2 & 2 & -1
\end{array}
\right) \, .
\ee
We note that the parametrisation of the matrix $m^{eff}_\nu$ in
Eq.(\ref{mnuLO}) also captures all corrections to the neutrino mass matrix 
which have been computed in the original model (without the field $\eta$) up
to a relative order $\la^4$ (with respect to the leading term), cf. Eq.(5.17)
in \cite{HKL10}. 
Defining the complex parameter
\be
{n=\frac{(-2a_\nu^*+b_\nu^*+c_\nu^*)d^{}_\nu + (b^{}_\nu-c^{}_\nu)d_\nu^*}{4 \left(\mathrm{Re}(b^{}_\nu c_\nu^* -a^{}_\nu  (b_\nu^*+c_\nu^*)) +  |a_\nu|^2  \right)} }\ , 
\ee
we can express the mixing angles of the neutrino sector as
\be
\label{thetanu}
\sin \theta_{13}^\nu \approx \frac{|n|}{\sqrt{2}}  \,\lambda   \ , \quad
\sin \theta_{23}^\nu\approx \frac{1}{\sqrt{2}} \left(1-  \frac{\mathrm{Re}(n)}{2} \la
\right) \  , \quad 
\sin \theta_{12}^\nu\approx \frac{1}{\sqrt{3}} \left( 1+\frac{|n|^2}{4} \lambda^2\right)  .
\ee
In order to obtain the lepton mixing we need to consider also the contributions from the
charged lepton sector. The latter are identical to those in the original model
\cite{HKL10} (see comments in sections \ref{recap} and
\ref{eta_charged_sector}) and are given in terms of the (positive) Yukawa
couplings $\tilde{x}_2$, $y_s$ and the phase $\alpha_{d,1}$. The resulting
lepton mixing angles read 
\be
 \sin \theta_{13}^l = \frac{r}{\sqrt{2}}  \; , \quad
\sin \theta_{23}^l = \frac{1}{\sqrt{2}} \, \left( 1+a \right) \; , \quad
\sin \theta_{12}^l = \frac{1}{\sqrt{3}} \, \left( 1+s \right) \, ,
 \ee
and the parameters $r$, $a$ and $s$ \cite{rsa} are given at LO in $\lambda$ by
\bea
&&
r\approx \left|n - \frac{\tilde x_2}{3y_s} \,  e^{-i \alpha_{d,1}}
\right|~\lambda \ , \qquad
a\approx - \frac{\mathrm{Re}(n)}{2} ~\lambda \ ,\qquad
s\approx -\frac{\tilde x_2}{3y_s} \, \cos{\alpha_{d,1}} ~\lambda \ . ~~~~~~~~~~~~\label{asr}
\eea
For the Jarlskog invariant $J_{CP}^l$ in the lepton sector (defined in the
same way as the one in the quark sector, see \cite{jcp}) we find
\bea
\label{eq:jcpl}
J_{CP}^l&\approx & 
- \frac 16 \left( \mathrm{Im}(n)
+ \frac{\tilde x_2}{3 y_s} \,  \sin{\alpha_{d,1}} \right) \, \lambda  \ .
\eea
As one can see, the first contribution arises from the neutrino sector, while
the second one is due to the charged lepton sector. In \cite{HKL10} only the
second term in Eq.(\ref{eq:jcpl}) is present (at this level in $\la$), because
$d_\nu \neq 0$ is only induced at the relative order $\la^4$.

Assuming that the value of the CP phase $\delta^l$ is dominated by the
contribution from the neutrino sector,\footnote{This assumption is reasonable
  taking into consideration that also the value of the mixing angle
  $\theta_{13}^l$ is dominated by the contribution from the neutrino sector.} 
we find $\sin\delta^l \approx -\mathrm{Im}(n)/|n|$. Using this and that
$\theta_C \approx \tilde{x}_2/ y_s \la$, see section \ref{recap}, we get 
\be
\label{abound}
|a| ~\lesssim~ \frac{1}{2} \left(r + \frac{\theta_C}{3}\right) |\cos\delta^l| \, .
\ee
For $r \approx 0.2$ (which implies $\theta_{13}^l \approx 0.15$) and CP
conservation $|\cos \delta^l|=1$ this bound on the deviation from maximal
atmospheric mixing translates into $0.38 \lesssim \sin^2 \theta_{23}^l
\lesssim 0.64$, which is comparable to the $3 \, \sigma$ range quoted
in \cite{fogli}. Obviously, for a non-trivial CP phase this bound 
will be tighter. For the prospects of measuring $\delta^l$ in the mid-term future, see for
example \cite{NOvA}. Similarly, we get for the deviation of the solar mixing
angle from its TB mixing value\footnote{Although the phase $\alpha_{d,1}$ is
  in principle determined by the CP violation in the quark sector,
  cf. Eq.(5.10) in \cite{HKL10}, we treat $\alpha_{d,1}$ here as free
  parameter, since, as has been shown in \cite{HKL10},  the correct amount of
  CP violation cannot be reproduced in this model, if $\alpha_{d,1}$  is
 the only source of CP violation.}
\be
\label{sbound}
|s|\lesssim \frac{\theta_C}{3} \  .
\ee
This translates into $0.29 \lesssim \sin^2 \theta_{12}^l \lesssim 0.38$. The
lower bound is tighter than the $3 \, \sigma$ bound found in \cite{fogli},
whereas the upper bound is a bit weaker than the bound achieved with a global
fit analysis of the experimental data. One should notice that these bounds are
given for mixing angles evaluated at a scale close to the scale of grand
unification, i.e. corrections coming from renormalisation group running have
not yet been included, and also contributions resulting, for example, from a
non-canonical K\"{a}hler potential have been neglected. As has been argued,
these effects can be small \cite{RGnoncanon}. We remark that identical sum
rule bounds also arise in the case of a SUSY $A_4 \times SU(5)$ model
\cite{Cooper:2012wf}. This is clear, since also in this model lepton mixing is
TB at LO, $\theta_{13}^l$ is generated in the neutrino sector by a suitable
breaking of one of the two $Z_2$ symmetries and the leading corrections from
the charged lepton sector are of the same form as in this model. 

Note that the contribution to the light neutrino mass matrix $m^{eff}_\nu$
originating from the Weinberg operator is at most of the relative order
$\la^8$ with respect to the leading contribution from the operators in
Eq.(\ref{LOnu}), if we assume the Weinberg operator to be suppressed by the
generic messenger scale $M$,  which is supposed to be of the order of the
scale of grand unification.\footnote{The largest contribution arises from $(F
  H_{5})^2 (\Phi^d_3)^2/M^3$ and breaks TB as well as  tri-maximal mixing.}

\subsection{Subleading operators involving ${\bs \eta}$}
\label{eta_charged_sector}

Since the additional flavon $\eta$ transforms as a trivial singlet under $S_4$, we
find the following operator (up to $\la^8$ assuming the sizes of the flavon
VEVs as given above) 
\be
\label{opcheta}
F T_3 H_{\ol{5}} \, \wt\Phi^d_3 \eta/M^2  \ ,
\ee
which contributes to the down quark and the charged lepton mass matrix. Using
the LO VEV of $\wt\Phi^d_3$, see Eq.(\ref{VEVdown}),   the operator leads to a
contribution to the (32) ((23)) and (33) entries of $M_d$ (and $M_e$) which is
of order $\la^7$. Such contributions can be absorbed into the parameters
present in the down quark and charged lepton mass matrices found in
\cite{HKL10} and thus do not change the results for fermion masses and mixing
presented there. Apart from the operator in Eq.(\ref{opcheta}) no additional
operators involving $\eta$ and contributing directly to the charged fermion mass
matrices up to the order $\la^8$ are generated. 


\section{Flavon superpotential}
\label{flavons}
\cleqn


A crucial ingredient for the construction of the flavon superpotential is the
assumption of a continuous $R$-symmetry under which matter superfields
carry charge $+1$, flavons and Higgs fields are uncharged and fields driving the alignment of the 
flavon VEVs have charge $+2$. As a consequence, the equations determining the
vacuum alignment are given by the $F$-terms of the driving fields which only
appear linearly in the superpotential. The latter also do not have any direct
couplings to matter superfields. 

\subsection{Impact of ${\bs \eta}$ on the vacuum alignment}

As one can check the field $\eta$ does not lead to any operator which strongly
perturbs the vacuum alignment of the flavons achieved with the help of the
driving fields $X^d_1$, $Y^d_2$, $Z^\nu_{3'} $, $Y^\nu_2$, $\ol X^d_1$,
$X^{\nu d}_{1'}$, $Y^{du}_2$ , $X^u_1$ and $X^{\mathrm{new}}_1$, $\wt
X^{\mathrm{new}}_{1'}$, found in table 3 and in Eq.(5.1) of \cite{HKL10}.  
At the subleading level the most important new operators are
\be
\label{etaflavonleading}
Y^\nu_2 \eta \Phi^d_2 \Phi^\nu_1/M + Y^\nu_2 \eta \Phi^d_2 \Phi^\nu_2/M +
Z^\nu_{3'} \eta \Phi^d_2 \Phi^\nu_{3'}/M \; ,
\ee
which are responsible for shifts in the vacuum of the fields $\Phi^\nu_2$ and
$\Phi^\nu_{3'}$ and thus give rise to additional contributions to the light
neutrino mass matrix which lead to further deviations from TB mixing. Using
the generic size of the flavon VEVs, these operators contribute at the level
$\la^9$, while the LO alignment of the fields $\Phi^\nu_2$ and $\Phi^\nu_{3'}$
originates from terms of the order $\la^8$ 
\be
Y^\nu_2 \Phi^\nu_1 \Phi^\nu_2 + Y^\nu_2 (\Phi^\nu_2)^2 + Y^\nu_2 (\Phi^\nu_{3'})^2
+  Z^\nu_{3'} \Phi^\nu_1 \Phi^\nu_{3'} + Z^\nu_{3'} \Phi^\nu_2 \Phi^\nu_{3'} + Z^\nu_{3'} (\Phi^\nu_{3'})^2 \, .
\ee
The conditions imposed by the $F$-terms of the fields $Y^\nu_{2}$ and
$Z^\nu_{3'}$ not only determine the alignment of the vacuum of the fields
$\Phi^\nu_1$, $\Phi^\nu_2$ and $\Phi^\nu_{3'}$, but they also relate the VEVs
$\varphi^\nu_i$ so that only one free parameters exists, cf. Eq.(4.8) in
\cite{HKL10}. We can thus parametrise the shifted VEVs as
\be
\langle\Phi^\nu_{3'} \rangle~=~  \begin{pmatrix} \varphi^\nu_{3'} +\Delta^\nu_{3',1} \\ \varphi^\nu_{3'}
+\Delta^\nu_{3',2} \\ \varphi^\nu_{3'} + \Delta^\nu_{3',3}\end{pmatrix}
\  , \qquad
\langle\Phi^\nu_2\rangle~=~ \begin{pmatrix} \varphi^\nu_2 +\Delta^\nu_{2,1}\\ \varphi^\nu_2 +\Delta^\nu_{2,2}\end{pmatrix}
\;\;\; \mbox{and} \;\;\;
\langle \Phi^\nu_1 \rangle~=~ \varphi^\nu_1  \ ,
\ee
with $\varphi^\nu_1$ being a free parameter. For the shifts in the VEVs of the
neutrino flavons we find  
\be
\label{nushifts}
\Delta^\nu_{2,1} \neq \Delta^\nu_{2,2} \;\;\; \mbox{and} \;\;\;  \Delta^\nu_{3',1}=\Delta^\nu_{3',2}=\Delta^\nu_{3',3}
\;\;\;\mbox{with}\;\;\; \Delta^\nu_{i,j}/M = \delta^\nu_{i,j} \la^5
\ee
still preserving the $Z_2$ symmetry generated by $S$ and thus tri-maximal
mixing in the neutrino sector. Most importantly, the alignment of the VEV of
$\Phi^\nu_{3'}$ is not disturbed. This is due to the fact that the operators
determining the shifts $\Delta^\nu_{i,j}$ have the following structures (up to
order $\la^{12}$, $D^\nu=Y^\nu_2, Z^\nu_{3'}$, $a, b=1,2,3'$):  
\be
D^\nu \Phi^\nu_a \Phi^\nu_b \;\; , \; D^\nu \eta \Phi^d_2 \Phi^\nu_a/M  \;\; , \; D^\nu \eta^2 (\Phi^d_2)^2/M^2 \;\; , \; 
D^\nu \Phi^\nu_a \Phi^\nu_b \wt\Phi^u_2/M \;\; , \; D^\nu (\Phi^d_2)^8 \Phi^\nu_a/M^7 \; . 
\ee
As one can check, they only involve flavons whose VEV alignment preserves the
generator $S$ and, hence, the shifts $\Delta^\nu_{i,j}$ are of the form in
Eq.(\ref{nushifts}).\footnote{See also \cite{HKL10} in which we have shown
  that $\Delta^\nu_{3',i}$ are equal up to a relative order of $\la^4$.} The
shifts $\Delta^\nu_{i,j}$ are enhanced by a factor $\la^{-3}$ compared to the
original model, cf. Eq.(5.3) in \cite{HKL10}. Thus, they contribute at the
same level, if plugged into the leading operators in Eq.(\ref{LOnu}), as the
operator in Eq.(\ref{theta13op}), to the neutrino mass matrix. In particular,
they also induce a non-zero value of $\theta_{13}^\nu$ in the neutrino
sector. Due to the preservation of the generator $S$ all such contributions
can be cast into the form of $m^{eff}_\nu$ shown in Eq.(\ref{mnuLO}). 
Furthermore, these enhanced shifts do not generate corrections to the charged
fermion mass matrices which are of order $\la^8$ or larger. 

There are several operators involving the driving fields and the field $\eta$
which arise at the level $\la^{11}$  or smaller and are irrelevant for the
discussion of the vacuum alignment as well as for the size of the leading VEV
shifts. Thus, we get the same results for the sizes of the VEV shifts as in
\cite{HKL10} apart from those of the flavons $\Phi^\nu_2$ and $\Phi^\nu_{3'}$,
as explained. 

\subsection{Relating the VEVs}

As discussed in one of the appendices of \cite{HKL10} it is possible to add
driving fields whose $F$-terms give rise to relations between different VEVs,
relate the latter to explicit mass scales as well as enforce the spontaneous
breaking of the flavour symmetry. In \cite{HKL10} two such fields $V_0 \sim
({\bf 1}, 0)$ and $V_2 \sim ({\bf 2}, -8)$ under $(S_4,U(1))$ have been
proposed. Adding the field $\eta$ does not disturb the relation induced by the
$F$-term of $V_0$, however it leads to problems with the field $V_2$. From the
$F$-terms of the latter one can derive two independent equations relating the
VEVs  $\varphi^{\nu}_1$, $\wt\varphi^u_2$,  $\varphi^d_2$ with an explicit
mass scale. The contribution to the $F$-term of $V_2$ coming from
\be
\label{V2dang}
V_2 \, \eta \Phi^d_2
\ee
is larger (it is of order $\la^5$) than the leading contribution without the
field $\eta$ (which is of order $\la^8$) and strongly perturbs the  results
achieved before.  In order to avoid this,  we consider instead the field $V_1$
which carries the same $U(1)$ charge as $V_2$, but is a trivial singlet under
$S_4$: $V_1 \sim ({\bf 1},-8)$.\footnote{Another possibility would be to
  consider the field $V_{3'} \sim ({\bf 3'},-8)$, because also this field
  cannot couple in an $S_4$-invariant way to the combination $\eta \Phi^d_2$
  and thus an operator similar to the one in Eq.(\ref{V2dang}) is
  forbidden. However, at the subleading level more operators arise than in the
  case of the field $V_1$.} Up to the order $\la^8$ the following terms can
contribute 
\be
M_{V_1} V_1 \Phi^\nu_1 + V_1 \wt\Phi^u_2 \Phi^\nu_2  + V_1 (\Phi^d_{2})^8/M^6 \, .
\ee
Plugging in the LO vacuum alignment of the flavons renders the last term
irrelevant and thus we get a correlation between $\wt\varphi^u_2$,
$\varphi^\nu_1$ and $M_{V_1}$. We can determine $\wt\varphi^u_2$ and it has
order $\la^4 M$, if we choose the mass scale $M_{V_1}$ to be of the order
$\la^4 M$. Corrections to the relation obtained from the $F$-term of $V_1$
arise from terms at order $\la^9$ and higher and are not relevant because they
only slightly change the value of the VEVs given in terms of parameters of the
flavon superpotential. 

\subsection{Fixing the VEV of ${\bs \eta}$}

The simplest possibility to relate the VEV of the field $\eta$ to an explicit
mass scale is to introduce a driving field $V_{\eta}$ transforming as ${\bf
  1}$ under $S_4$ and carrying $U(1)$ charge $-7$. Then the term 
\be
\label{VetaLO}
M_{\eta} V_\eta \eta
\ee
is allowed. Furthermore, up to order $\la^9$ the structures
\be
V_\eta (\Phi^d_2)^7/M^5 + V_\eta (\Phi^d_2)^3 \Phi^d_3 \Phi^\nu_{3'}/M^3
\ee
are compatible with all symmetries of the model. Plugging in the LO vacuum,
only the latter is relevant and leads together with the operator in
Eq.(\ref{VetaLO}) to a relation between the VEV of $\eta$ and $\varphi^d_2$,
$\varphi^d_3$ and $\varphi^\nu_1$. Choosing $M_\eta \sim \la^5 M$ gives rise
to $\langle \eta \rangle \approx \la^4 M$ as required for having
$\theta_{13}^l$ of order $\la$.  

When recomputing the shifts in the VEVs of the flavons considering a
superpotential with the fields $V_0$, $V_1$ and $V_\eta$, we have to take into
account three additional shifts $\wt\Delta^u_{2,2}$, $\Delta^\nu_1$ and
$\Delta^\eta$ which are related to the three additional equations coming from
the $F$-terms of $V_0$, $V_1$ and $V_\eta$. These do not lead  to new
contributions to the fermion mass matrices which cannot be absorbed into the
existing ones, but are relevant for consistently solving the $F$-term
equations. We note that we find an enhancement by $\la^{-1}$ of the shifts
$\wt\Delta^d_{3,2}$  and $\wt\Delta^d_{3,3}$ in the VEVs of the second and
third components of the field $\wt\Phi^d_3$ and of the shift $\Delta^u_{2,2}$
in the VEV of the second component of $\Phi^u_2$ 
\be
\label{shiftenhance}
\wt\Delta^d_{3,2}/M = \wt\delta^d_{3,2} \, \la^4 \; , \;\; \wt\Delta^d_{3,3}/M
= \wt\delta^d_{3,3} \, \la^4  \;\;\; \mbox{and} \;\;\; \Delta^u_{2,2}/M =
\delta^u_{2,2} \, \la^5 \ ,
\ee
compare to Eq.(5.3) in \cite{HKL10}. However, the shifts $\wt\Delta^d_{3,2}$
and $\wt\Delta^d_{3,3}$ are equal at this level and can thus be absorbed into
the VEV $\wt\varphi^d_3$. The relevant difference between the VEVs of the two
components $\wt\Phi^d_{3,2}$ and $\wt\Phi^d_{3,3}$ still arises at the order
$\la^5$ and the results of fermion masses and mixing are not
altered. Similarly, the effect of the enhanced shift $\Delta^u_{2,2}$ is not
relevant, because its effect can be absorbed into the LO VEV $\varphi^u_2$.    
\begin{table}
\begin{center}
\begin{tabular}{|c||c|c|c|c|c|c|c|c|}
\hline
\rule[0.16in]{0cm}{0cm} \!Field\phantom{\Big|}\!\!\! & $\ol{B}$ & $B$ & $A$ & $\ol{A}$ & $\Xi$ & $\ol{\Xi}$&$\Gamma$&$\ol\Gamma$\\
\hline
\rule[0.16in]{0cm}{0cm}$SU(5)$\phantom{\Big|}\!\! & $\bf \overline{5}$ & $\bf 5$ &  $\bf 10$ & $\bf \overline{10}$ & $\bf 10$ & $\bf \overline{10}$&${\bf 1}$&${\bf 1}$\\
\hline
$S_4$\phantom{\Big|}\!\! & $\bf 1$ & $\bf 1$ & $\bf 2$ & $\bf 2$ & $(\bf{1},\bf{1'})$ & $(\bf{1},\bf{1'})$& $\bf{3}$& $\bf{3}$\\
\hline
$U(1)$\phantom{\Big|}\!\! & $0$ & $0$ & $-5$ & $5$ & $5$ & $-5$&$-3$&$3$\\
\hline
\end{tabular}\end{center}
\begin{center}
\caption[]{Heavy fields which render the operators given in Eqs.(\ref{LOup},
  \ref{theta13op}) and the first operator in Eq.(\ref{LOdown}) renormalisable. 
Note that $\Xi$ and $\ol{\Xi}$ can either transform as ${\bf 1}$ or ${\bf 1'}$
under $S_4$ and none of the two choices is preferred over the other. All
fields carry a $U(1)_R$ charge $+1$. \label{heavyfields_new}} 
\end{center}
\end{table}
%


\section{Messengers}
\label{messengers}
\cleqn


We present here a set of messengers which allows to UV-complete all leading
operators contributing directly to fermion mass matrices and not only the last
two ones mentioned in Eq.(\ref{LOdown}). In doing so we restrict ourselves to
consider matter messengers only which have $U(1)_R$ charge $+1$ like the
superfields $T$, $T_3$, $F$ and $N$.  

As has been discussed in detail in \cite{HKL10} a viable UV completion leading
to the last two operators in Eq.(\ref{LOdown}) is given by the messengers:
$\Sigma \sim ({\bf 10},{\bf 1}, 6)$, $\Delta \sim ({\bf 5},{\bf 1}, 7)$,
$\Upsilon \sim ({\bf 5},{\bf 2}, 5)$, $\Omega \sim ({\bf 5},{\bf 3}, -6)$,
$\Theta \sim ({\bf 5},{\bf 3}, -5)$ under $(SU(5),S_4,U(1))$ and their
vector-like partners\footnote{Note that we have slightly changed notation with
  respect to \cite{HKL10}. However, the quantum numbers of the messengers are
  the same and thus they lead to the same terms in the superpotential.} giving
rise to the diagrams shown in figure 1 of \cite{HKL10}. 

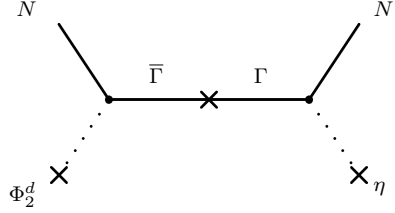
\begin{figure}[t!]
\begin{center}
\begin{fmffile}{diagrameta}
  \begin{fmfgraph*}(50,20)
    \fmfleft{i1,i2}
    \fmfright{o1,o2}
    \fmf{dots}{v1,i1}
    \fmflabel{$\mbox{\scriptsize $\Phi^d_2$ \footnotesize}$}{i1}
    \fmfv{decor.shape=cross,decor.filled=full,decor.size=4thick}{i1}
    \fmf{plain}{v1,i2}
    \fmfv{decor.shape=circle,decor.filled=full,decor.size=1thick}{v1}
    \fmflabel{$\mbox{\scriptsize $N$ \footnotesize}$}{i2}
    \fmf{plain,label=$\mbox{\scriptsize $\ol{\Gamma}$ \footnotesize}$}{v1,v2}
    \fmfv{decor.shape=cross,decor.filled=full,decor.size=4thick}{v2}
    \fmf{plain,label=$\mbox{\scriptsize $\phantom{\ol{\Gamma}}\Gamma$ \footnotesize}$}{v2,v3}
    \fmfv{decor.shape=circle,decor.filled=full,decor.size=1thick}{v3}
    \fmf{dots}{v3,o1}
    \fmflabel{$\mbox{\scriptsize $\eta$ \footnotesize}$}{o1}
    \fmfv{decor.shape=cross,decor.filled=full,decor.size=4thick}{o1}
    \fmf{plain}{v3,o2}
    \fmflabel{$\mbox{\scriptsize $N$ \footnotesize}$}{o2}
   \end{fmfgraph*}
\end{fmffile}
\end{center}
\begin{center}
\caption{Diagram for generating the contribution coming from the operator
  $\eta \Phi^d_2 N N/M$ in a renormalisable theory. Scalars/fermions are
  displayed by dotted/solid lines. Crosses indicate a VEV for scalar
  components and mass insertions for fermions. 
\label{diagrameta}}
\end{center}
\end{figure}

In order to UV-complete the operator generating the mass of the bottom quark
and the tau lepton, see Eq.(\ref{LOdown}), we add the messengers $B$ and
$\ol{B}$ with quantum numbers as found in table \ref{heavyfields_new}:
\be
\kappa_1 T_3 \ol{B} H_{\ol{5}} + \kappa_2 B F \Phi^d_3 + M_B \ol{B} B  \, .
\ee
The effective operators giving rise to up and charm quark masses, see
Eq.(\ref{LOup}), are promoted to renormalisable operators with the help of the
messengers $A$ and $\Xi$ and their vector-like partners\footnote{Note that the
  up quark mass also receives a contribution from the operator dominantly
  generating the charm quark mass, if the shifted vacuum of $\Phi^u_2$
  ($\Delta^u_{2,1}/M=\delta^u_{2,1} \, \la^8$) is considered. Thus, the
  presence of the messenger $\Xi$ is strictly speaking not necessary in order
  to generate the mass of the up quark of the correct order of
  magnitude. However, in order to match the operators present in the effective
  theory we have included this messenger.} 
\be
\rho_1 T A H_5 + \rho_2 T \ol{A} \Phi^u_2  
+\lambda_1 \ol{A} \Xi \Phi^u_2 + \lambda_2 T \ol{\Xi} \wt\Phi^u_2 + M_A A \ol{A} + \gamma_4 A \ol{A} \wt\Phi^u_2 + M_{\Xi} \Xi \ol{\Xi} \, .
\ee
The operator, responsible for the largish value of the lepton mixing angle
$\theta_{13}^l$, see Eq.(\ref{theta13op}), arises in the UV completion from
\be
\sigma_1 N \ol{\Gamma} \Phi^d_2 + \sigma_2 N \Gamma \eta + M_\Gamma \ol{\Gamma} \Gamma +\gamma_5 \ol{\Gamma} \Gamma \wt\Phi^u_2 \, .
\ee
The diagram belonging to these messengers can be found in figure \ref{diagrameta}.
%
\setlength{\unitlength}{1mm}
%
Integrating out the messengers, present in table \ref{heavyfields_new},  and
plugging in the LO vacuum alignment, see Eqs.(\ref{VEVup}-\ref{VEVnu}), the
contributions to the fermion mass matrices read (up to order $\la^8$) 
\bea
&&-\kappa_1 \kappa_2 \left(\varphi^d_3/M_B \right) F_3 T_3 H_{\ol{5}} \; ,\\
&& \lambda_1 \lambda_2 \rho_1 \left( \varphi^u_2\wt\varphi^u_2/(M_A M_\Xi) \right) T_1 T_1 H_5 - \rho_1 \rho_2 \left( \varphi^u_2/M_A \right) T_2 T_2 H_5 \; ,\\
&& -\sigma_1 \sigma_2 \left( \langle\eta\rangle \varphi^d_2/M_\Gamma \right) \left( N_1 N_2 + N_2 N_1 + N_3 N_3 \right) \; .
\eea
Contributions coming from, for example, the presence of the coupling
$\gamma_4$ turn out to be of order $\la^9$ or smaller and thus are irrelevant,
while contributions associated with the shifts in  the flavon VEVs are of the
same form and order as in the effective theory, see for details \cite{HKL10}. 

We find one additional term
\be
\label{termadd}
x \ol{\Delta} B \eta
\ee
which gives rise to the operator in Eq.(\ref{opcheta}). The latter induces a
contribution of order $\la^7$ to the down quark and charged lepton mass matrix
which is of the form 
\be
\alpha_3 \kappa_1 x \left( \wt\varphi^d_3 \langle\eta\rangle/(M_B M_\Delta)\right) \, \left( F_2 T_3 - F_3 T_3 \right) H_{\ol{5}}\, ,
\ee
if the messengers are integrated out and the LO VEVs are plugged in. 
We note that the coupling $\alpha_3$ is defined in Eq.(B.1) of \cite{HKL10} as the coupling of
the operator $\Delta F \wt{\Phi}^d_3$. 


\section{Alternative extension with ${\bs \eta \sim \bf{1'}}$}
\label{eta1pr}
\cleqn


In the above extension we have assumed that $\eta \sim {\bf 1}$ under $S_4$.
If we assume instead $\eta \sim {\bf 1'}$ under $S_4$, we can still write down
the operator in Eq.(\ref{theta13op}) which is crucial for
generating~$\theta_{13}^l$. Again, we preserve the $Z_2$ symmetry generated by 
$S$, because also the singlet ${\bf 1'}$ has $S=1$. The generator $U$, on the
other hand, is now broken by the VEVs of both fields $\Phi^d_2$ and $\eta$,
since $U=-1$ for the representation ${\bf 1'}$. Several of the subleading
operators differ. The operator in Eq.(\ref{opcheta}) which contributes at the
order $\la^7$ to the down quark and charged lepton mass matrix is not
invariant under $S_4$ for $\eta\sim {\bf 1'}$ and thus absent. In the flavon
superpotential the most relevant operators, see Eq.(\ref{etaflavonleading}),
exist independently of the choice of $\eta \sim {\bf 1}$ or $\eta \sim {\bf
  1'}$ and lead to the same results for the leading shifts  $\Delta^\nu_{i,j}$
in the VEVs of the flavons $\Phi^\nu_1$, $\Phi^\nu_2$ and $\Phi^\nu_{3'}$. In
addition, an operator exists at the subleading level $\la^{10}$ which is of the form 
 \be
  \wt X^{{\rm new}}_{1'} \eta (\wt\Phi^d_3)^2/M \, .
\ee
Its effect is to enhance the shifts $\wt\Delta^d_{3,2}$, $\wt\Delta^d_{3,3}$
and $\Delta^u_{2,2}$ in the same way as in the case of $\eta\sim {\bf 1}$, if
we consider a superpotential containing the field $V_0$, $V_1$ and $V_\eta$,
see Eq.(\ref{shiftenhance}). Again, these shifts are irrelevant because they
do not change the results of fermion masses and mixing. The discussion
concerning the driving fields $V_0$, $V_2$ and $V_1$ holds for $\eta \sim {\bf
  1'}$ as well. In order to relate the VEV of $\eta\sim {\bf 1'}$ to those of
the other flavons and mass scales in the model we use a field $V_\eta$ which
now transforms as ${\bf 1'}$. Eventually, the transformation properties of the
messengers $\Gamma$ and $\ol\Gamma$ depend also on the nature of the field
$\eta$: if we choose $\eta\sim {\bf 1'}$, the messengers have to transform as
${\bf 3'}$. Note that the operator in Eq.(\ref{termadd}) is not
$S_4$-invariant for $\eta \sim {\bf 1'}$ and thus no such subleading
contribution to the down quark and charged lepton mass matrices is present in
this case. This is consistent with our findings in the effective theory that
the operator in Eq.(\ref{opcheta}) is forbidden for $\eta\sim {\bf 1'}$. 


\section{Conclusions}
\label{conclusions}
\cleqn


We have discussed a simple extension of an existing SUSY $S_4 \times 
SU(5)$ model \cite{HKL10} which, in its original form, has been ruled out by 
the recent measurements of $\theta_{13}^l \approx 0.15\div 0.17$ in the Daya 
Bay and RENO experiments. We have shown how augmenting the model with only one
additional $S_4$ singlet flavon gives rise to a sizable $\theta^l_{13}$ via 
an operator contributing to the neutrino mass matrix. This contribution,
 which is suppressed by $\la$ relative to the leading terms, breaks one of the two 
$Z_2$ symmetries preserved in the neutrino sector at LO, namely the one
 generated by the element $U$. It is identified with $\mu - \tau$ symmetry and
 is responsible for a vanishing neutrino mixing angle $\theta_{13}^\nu$; the
 second $Z_2$ symmetry, associated with the generator $S$, remains intact and
 enforces tri-maximal (neutrino) mixing. The successful predictions of fermion
 masses, quark mixing and the solar mixing angle achieved in the original
 model are maintained. The corrections to the atmospheric mixing angle are
 enhanced due to the breaking of the symmetry generated by $U$ at relative
 order $\la$. The deviations of the solar and the atmospheric mixing angle
 from their TB mixing values are subject to simple sum rule bounds, see
 Eqs.(\ref{abound},\ref{sbound}).  Finally, we have presented a simple UV
completion of the operators directly relevant for fermion masses and mixing. 
It might be interesting to revisit other models which predict a specific
mixing pattern with $\theta_{13}^l=0$ or $\theta_{13}^l$ too small to
accommodate the recent results of Daya Bay and RENO. A modest extension of the
particle content, analogous to the one discussed in this note, might induce a
suitable breaking of the symmetry that is responsible for the smallness of the
mixing angle $\theta_{13}^l$ without spoiling the successful predictions of
the model. 


\section*{Acknowledgements}


We have been partly supported by the European Programmes
PITN-GA-2009-237920-UNILHC and PITN-GA-2011-289442-INVISIBLES. SFK also
acknowledges partial support from the STFC Consolidated ST/J000396/1. CH would
like to thank the Aspen Center for Physics for kind hospitality at the initial
stage of this work. 


\end{document}